\documentclass{article}
\usepackage{spconf,amsmath,graphicx}
\usepackage{url}
\usepackage{subfigure}
\usepackage{caption}
\usepackage{booktabs}
\usepackage{bbding}
\usepackage{array,multirow}
\title{Half Wavelet Attention on M-Net+ for Low-Light Image Enhancement}
\name{Chi-Mao Fan$^{\star}$ \qquad Tsung-Jung Liu$^{\star}$ \qquad Kuan-Hsien Liu$^{\dagger}$}
\address{
\ninept $^{\star}$Department of Electrical Engineering and Graduate Institute of Communication Engineering, National Chung Hsing University, Taiwan
\\
\ninept $^{\dagger}$Department of Computer Science and Information Engineering, National Taichung University of Science and Technology, Taiwan
}
\begin{document}

%\ninept
%
\maketitle

\begin{abstract}
Low-Light Image Enhancement is a computer vision task which intensifies the dark images to appropriate brightness. It can also be seen as an ill-posed problem in image restoration domain. With the success of deep neural networks, the convolutional neural networks surpass the traditional algorithm-based methods and become the mainstream in the computer vision area. To advance the performance of enhancement algorithms, we propose an image enhancement network (HWMNet) based on an improved hierarchical model: M-Net+. Specifically, we use a half wavelet attention block on M-Net+ to enrich the features from wavelet domain. Furthermore, our HWMNet has competitive performance results on two image enhancement datasets in terms of quantitative metrics and visual quality. The source code and pretrained model are available at \url{https://github.com/FanChiMao/HWMNet.}
\end{abstract}

%

% --section-- of Introduction
\section{Introduction}
Image enhancement is a low-level computer vision task which also plays an important role in the pre-process of high-level vision tasks such as image classification and object detection. Generally, image enhancement can also be considered as one kind of image restoration task, in which a low-quality image $Y$ could be represented as: 
\begin{equation}
Y = D(X) + n, \label{eq1}
\end{equation}
where $X$ is a high-quality or noise-free image, $D(\cdot)$ denotes the degradation function and $n$ means the additive noise. The image enhancement methods could be roughly classified to two types: algorithm-based and learning-based. Algorithm-based enhancement methods could also be summarized to three mainstream categories (i.e., histogram equalization-based, retinex-based, and dehazing-based methods). All of above methods are based on image priors which also could be called prior-based or model-based methods. More details are going to be discussed in the section of related work.

Recently, learning-based enhancement methods surpass conventional prior-based methods in terms of inference time and enhancement performance. Especially the appearance of convolution neural network (CNN) almost dominates the whole computer vision area, including image enhancement.

%, we try to design an extraction block which could extract the adequate and useful spatial feature information, and also trade the computation time off against the enhancement performance of the network structure. First
In this paper, we propose a hierarchical image enhancement model based on the existing M-Net architecture called M-Net+ to ameliorate the loss of spatial information caused from the sampling. Moreover, we use the proposed half wavelet attention block (HWAB) in our M-Net+ to gain the feature information in wavelet domain. For the reconstruction process, instead of using concatenation to fuse the feature maps with different resolutions, we adopt the selective kernel feature fusion (SKFF) \cite{01} to efficiently combine the features. Overall, the main contributions of this work can be summarized as follows: 

\begin{itemize}
  \item We propose the improved hierarchical architecture (M-Net+) for the task of low-light image enhancement.
  \item We propose the efficient feature extraction block called half wavelet attention block (HWAB) which can gain more diversified features in another domain.
%  \item We propose the novel feature fusion based on wavelet domain called selective wavelet feature fusion (SWFF) which can reconstruct enhanced images with more details.
  \item We experiment on two low-light image enhancement datasets to demonstrate that our proposed model achieves state-of-the-art performance in image enhancement in terms of quantity and quality and even with less computational complexity.
\end{itemize}

% --section-- of Related work
\section{Related Work}
%In this section, we first are going to describe the development of several image enhancement methods. Then, we will discuss the balance between spatial and contextual information in the hierarchical architecture. Moreover, we will introduce the special hierarchical network architecture which is modified from the traditional U-Net \cite{unet}. Finally, we will illustrate the selective kernel which we apply in our proposed feature fusion method.

% --figure-- of HWMNet 
\begin{figure*}[!t]
\centering
	\includegraphics[width=17.5cm]{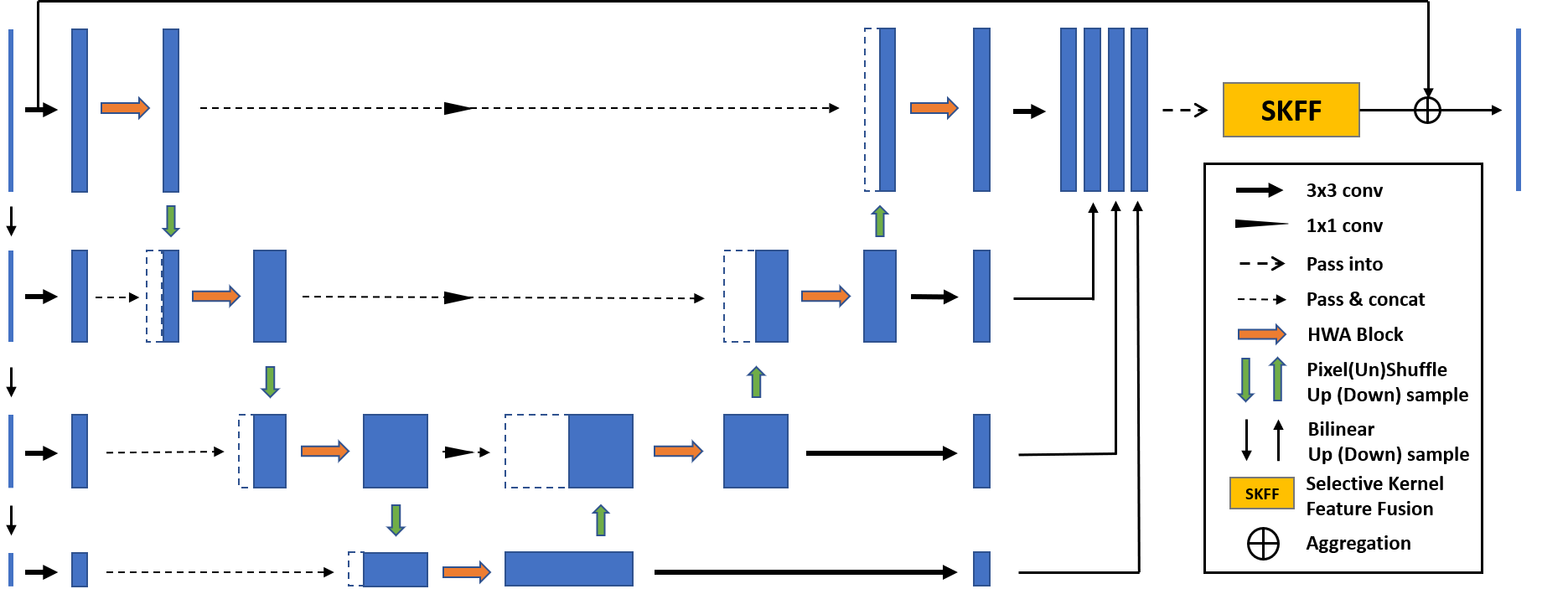}
	\caption{Proposed Half Wavelet attention M-Net (HWMNet) architecture. The source code and component structure of the model could be found in the provided URL indicated in the abstract. We set the initial channel in each resolution to 96 after $3 \times 3$ convolutions, and totally we have 4 layers in the proposed M-Net.}
	\label{HWMNet}
\end{figure*}

\subsection{Image enhancement}
%, which makes restored images less vivid
As aforementioned, traditional image enhancement approaches are generally based on image priors or algorithms, such as histogram equalization-based \cite{adaptive, contrast}, retinex-based \cite{MSR} and dehazing-based \cite{DCP} methods. The enhancement performances of conventional prior-based methods are acceptable but always suffered from color distortions or produce over-enhanced images. Fortunately, the learning-based methods based on CNNs recently have been successfully applied to low-light image enhancement task \cite{2019ntire}. And they almost surpass the conventional prior-based enhancement methods \cite{ 01, kind, kind++, RUAS, LLFlow} with incredibly good performance.

%\subsection{Hierarchical architecture} % 感覺太多把這段去掉，教授再看空間夠不夠，考慮要不要加回來
%For the image enhancement tasks, or more broadly, the image restoration tasks, the restoration models could be divided into two categories according to the model structure. One is single-scale architectures, the other is multi-scale networks architecture \cite{mnet, unet, fpd, LLFlow} or called by hierarchical architecture. The single-scale restoration model is good at keeping the spatial information which the restored images are more accurate since it is without any sampling operation. The image restoration tasks are usually be categorized to pixel-wise vision tasks, which means the position of pixels are important. However, single-scale model can preserve the information of pixel's location but the sematic information is not enriched. On the contrary, hierarchical network architectures have the ability to gain the adequate and plentiful sematic information but the spatial details will loss in the sampling process. So, the trade off between the spatial loss and enriched sematic feature information is worth considering.

\subsection{M-Net}
The M-Net architecture is first proposed for medical image segmentation \cite{mnet} which can be regarded as an improved hierarchical model architecture from U-Net \cite{unet}. Adiga \emph{et al.}\cite{fpd} use the same framework for fingerprint image denoising and also obtain good results. Compared with the traditional U-Net architecture, the M-Net adds additional gatepost feature paths in both encoder and decoder which the main objective is to enrich the contextual information in different image resolutions and try to ameliorate the spatial loss caused by sampling.

%\subsection{Selective Kernel Network}
%Li \emph{et al.}\cite{sknet} proposed the selective kernel convolution that has two branches. One of the paths utilizes normal naive $3 \times 3$ convolution kernel to extract features, and the other path adopts different kernel size (e.g. $5 \times 5$, $7 \times 7$) to obtain the larger receptive field. At the end of selective kernel convolution, they use softmax activation function to acquire the weights for two different features maps. Zamir \emph{et al.}\cite{01} were inspired by \cite{sknet} and applied it to multi-scale feature fusion for image enhancement tasks.

% --section-- of Proposed Method
\section{Proposed Method}
%In this section, we mainly introduce the proposed Half Wavelet Attention on M-Net+ (HWMNet), and provide detailed explanations for each ingredient of the model in the following subsections.

\subsection{Half Wavelet attention M-Net (HWMNet)}
The proposed architecture of HWNNet for low-light image enhancement is shown in Fig. \ref{HWMNet}. We first use $3 \times 3$ weight sharing convolution in each resolution of low-light input image acquired by doing the bilinear down-sampling from original-resolution input. Each layer has the proposed half wavelet attention block (HWAB) to extract adequate and enriched semantic information. More details about HWAB will be illustrated in Section 3.3. In addition, we use pixel unshuffled \cite{RDN} method as our down-sampling module in the trunk feature path at the end of HWAB in the U-Net. Then, the feature maps are concatenated with previous shallow features from bilinear down-sampling and keep going as normal U-Net process. In this paper, we name this hierarchical network as M-Net+ and it will be discussed in section 3.2. 

We follow the recent enhancement method \cite{01}, and optimize our HWMNet end-to-end with the Charbonnier loss which is a variant L1 loss for low-light image enhancement and described as follows: 

\newcommand{\Lagr}{\mathcal{L}} 
\begin{equation}
\begin{aligned}
\Lagr_{char} = \sqrt{ ||\hat{X} - X||^2 + \varepsilon^2}, \label{eq2}
\end{aligned}
\end{equation}
where $\hat{X}, X \in {R}^{N \times C \times H \times W}$ denote the low-light images and ground-truth images, respectively. $N$ is the batch size of training data, $C$ is the number of feature channels, $H$ and $W$ are the size of images. The constant $\varepsilon$ in Eq. (\ref{eq2}) is empirically set to $10^{-3}$.

%% --figure-- of Resize module and HWAB
%\begin{figure*}[!htbp] 
%\centering
	%\subfigure[]{
	%\begin{minipage}[t]{0.49\linewidth}
	%\centering
	%\includegraphics[width=8.7cm]{./figures/model/Resize_frame.png}
	%\label{Resize}
	%\end{minipage}%
	%}%
%\hfill
%\centering
	%\subfigure[]{
	%\begin{minipage}[t]{0.49\linewidth}
	%\centering
	%\includegraphics[width=8.7cm]{./figures/model/HWAB_frame.png}
	%\label{HWAB}
	%\end{minipage}%
	%}%
%\centering
%\caption{(a) Resizing module with pixel (un)shuffle. (b) Illustration of the Half Wavelet Attention Block (HWAB) in our HWMNet.}
%\label{block}
%\end{figure*}

\subsection{M-Net+}
After analyzing the conventional M-Net \cite{mnet, fpd} that has been discussed in previous section, we think there are two major problems which make M-Net not a popular hierarchical network backbone in recent image restoration methods. First, for the encoder, they use max-pooling as down-sampling in both gatepost and U-Net feature paths. Since the restoration tasks are the pixel-wise vision tasks, it is inappropriate to use the down-sampling which will cause serious loss of spatial information. Second, for the decoder, because the additional gatepost feature paths are added, the reconstruction process (decoder) has a lot of features to fuse, which will increase the burden to the whole network. 

To focus on these two problems, we proposed the M-Net+ as our enhancement network backbone. The proposed HWMNet has two improvements: 1) More diversity and plentiful multi-scale cascading features. We utilize pixel unshuffle down-sampling in the U-Net path and bilinear down-sampling for the gatepost path, which makes the cascading features have more diversity. Furthermore, the design of the gatepost feature paths in the M-Net can ameliorate the loss of spatial information; 2) Using different feature fusion method SKFF \cite{01} to summarize the information in the decoder (reconstruction process). This improvement aims to solve high-dimensional cascading features in original M-Net, which makes the M-Net have large number of parameters and high computational complexity. 

\subsection{Half Wavelet Attention Block}
\begin{figure}[!htbp] 
\centering
	\includegraphics[width=8.7cm]{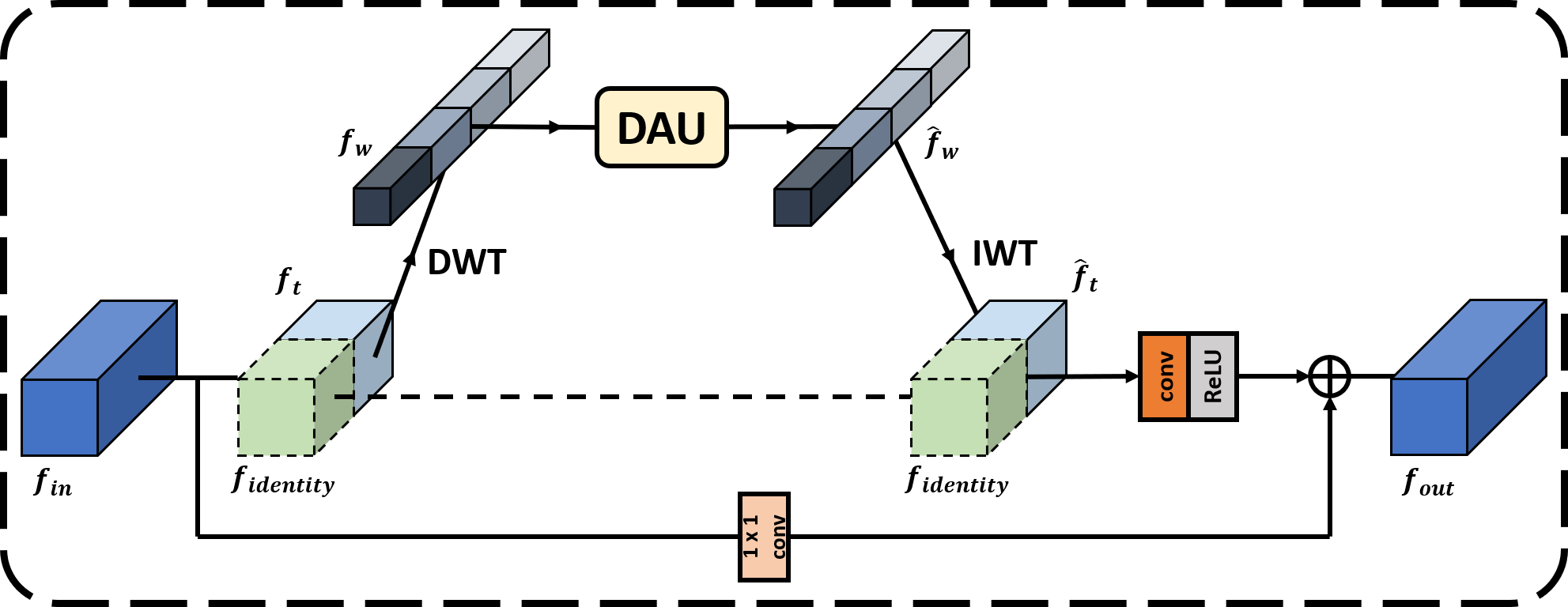}
\centering
\caption{Illustration of the Half Wavelet Attention Block (HWAB) in our HWMNet.}
\label{HWAB}
\end{figure}

Fig.~\ref{HWAB} shows the architecture of the proposed Half Wavelet Attention Block (HWAB) which is modified from the dual-attention unit (DAU) \cite{01, cycleisp}. The DAU uses channel attention \cite{SENet} and spatial attention to extract features, and fuses them together to gain more enriched contextual information.
%Due to the page limits, we don't display the architecture's schematic of the DAU and briefly show in Fig.~\ref{HWAB}. 

In the beginning of the HWAB, the input features $f_{in} \in {\mathbf{R}}^{C \times H \times W}$ are divided into two parts $f_{identity}$ and $f_{t}$ along the input channel, and they both $\in {\mathbf{R}}^{\frac{C}{2} \times H \times W}$. The main purpose of dividing the input features is to decrease the computation complexity and reserve the context information, where $f_{identity}$ is to keep the normal domain features which can be referred to HINet \cite{hinet}, while the other part of feature $f_{t}$ will do the discrete wavelet transformation (DWT) to obtain the wavelet domain feature $f_{w} \in {{\mathbf{R}}^{2C \times \frac{H}{2} \times \frac{W}{2}}}$. Then, $f_{w}$ will be passed to the DAU \cite{01, cycleisp} to get the weighted wavelet feature ${\hat{f}_{w} \in {{\mathbf{R}}^{2C \times \frac{H}{2} \times \frac{W}{2}}}}$ and we perform the inverse wavelet transformation (IWT) to reshape $\hat{f}_{w}$ to the shape of $f_{t}$ and become the weighted normal domain feature $\hat{f}_{t} \in {\mathbf{R}}^{\frac{C}{2} \times H \times W}$. After concatenating these two features ($\hat{f}_{t}$ and $f_{identity}$), the residual feature $f_r$ are obtained by passing these feature to $3 \times 3$ convolution layer and parameter ReLU layer. Finally, adding the shortcut features through $1 \times 1$ convolution to the residual feature $f_r$, we get the output feature $f_{out} \in {\mathbf{R}}^{C \times H \times W}$ which has the wavelet attention feature information.

% --figure-- HWAB

%\subsection{Selective Wavelet Feature Fusion}
%In this subsection, we are going to illustrate the details of proposed selective wavelet feature fusion (SWFF). The architecture of SWFF is showed in Fig.~\ref{SWFF}. Using one sentence to describe the proposed SWFF, we would be said "Aggregate each feature with weights which is focusing on wavelet domain". The features ($f_{1}, f_{2} \in {\mathbf{R}}^{C \times H \times W}$ in Fig.~\ref{SWFF}) denote as the to fuse features. They are doing the DWT and simply add together to acquire the wavelet domain feature $f_w \in {\mathbf{R}}^{4C \times \frac{H}{2} \times \frac{W}{2}}$ firstly. Then, these two input features $f_{1}, f_{2}$ will do the SWFF by multiplying the corresponding feature descriptor vectors ($v_{1}, v_{2} \in {\mathbf{R}}^{C \times 1 \times 1}$ which are generated from channel-wise statistics $s \in {\mathbf{R}}^{4C \times 1 \times 1}$) to get the weighted wavelet features ($\hat{f}_{1}, \hat{f}_{2} \in {\mathbf{R}}^{C \times H \times W}$). In the end of selective wavelet feature fusion, we aggregate two channel-weighted features $\hat{f}_{1}, \hat{f}_{2}$ to get the output fusion feature. The implementation code of SWFF could be found in the provided URL indicated in the abstract.

% --figure-- of SWFF 
%\begin{figure}[!t]
%\centering
%	\includegraphics[width=8.5cm]{./figures/model/SWFF.png}
%	\caption{The architecture of Selective Wavelet Feature Fusion.}
%	\label{SWFF}
%\end{figure}

% --section-- of Experiment
\section{Experiment}
\subsection{Experiment Setup}
\noindent\textbf{Implementation Details.} Our HWMNet is an end-to-end model and trained from scratch. The experiments conducted in this paper are implemented by PyTorch 1.8.0 with single NVIDIA GTX 1080Ti GPU. The network is trained on $256 \times 256$ patches only with a batch size of 2 for $1 \times{10}^5$ iterations. For data augmentation, randomly horizontal and vertical flips are adopted. We use Adam optimizer with the initial learning rate of $1 \times {10}^{-4}$ and decrease to $1 \times {10}^{-6}$ by the consine annealing strategy. 
%We also apply the warm-up strategy increasing the learning rate from $\frac{1}{3} \times {10}^{-4}$ to $1 \times {10}^{-4}$ in the first three epochs of training.

\noindent\textbf{Evaluation Metrics.} For the quantitative comparisons, we consider the Peak Signal-to-Noise Ratio (PSNR), Structural Similarity (SSIM) index and Learned Perceptual Image Patch Similarity (LPIPS) metric \cite{LPIPS}. Noted that the performance is better when both PSNR and SSIM values are higher. On the contrary, the LPIPS value is lower when the restoration performance is closer to human perception.

% --figure-- of visual result (LOL)
\begin{figure*}[!t] %\footnotesize
\centering
	\subfigure{
	\begin{minipage}{2.9cm}
	\includegraphics[width=2.8cm]{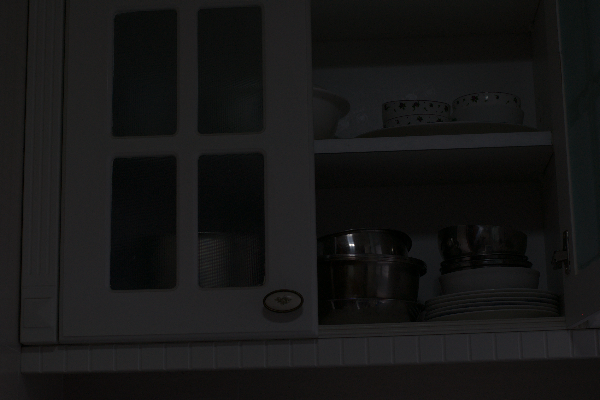}
	\vspace*{-2.5mm}
	\caption*{\small{5.09 / 0.197 / 0.523}}
	\vspace*{-4mm}
	\caption*{\small{Low-light}}
	\end{minipage}
	\begin{minipage}{2.9cm}
	\includegraphics[width=2.8cm]{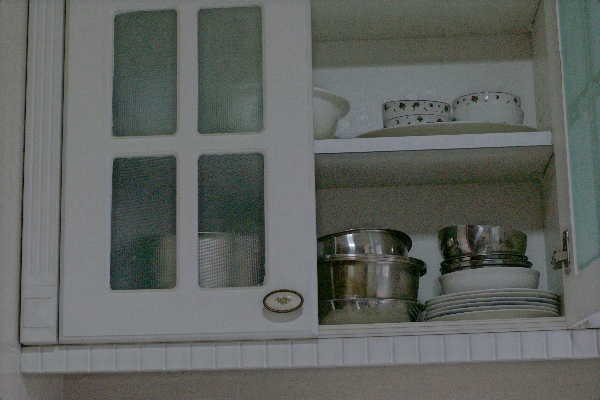}
	\vspace*{-2.5mm}
	\caption*{\small{11.87 / 0.637 / 0.291}}
	\vspace*{-4mm}
	\caption*{\small{Zero-DCE \cite{zeroDCE}}}
	\end{minipage}
	\begin{minipage}{2.9cm}
	\includegraphics[width=2.8cm]{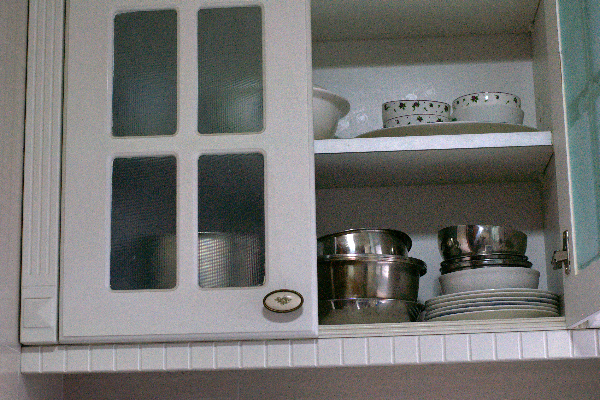}
	\vspace*{-2.5mm}
	\caption*{\small{14.68 / 0.576 / 0.286}}
	\vspace*{-4mm}
	\caption*{\small{LIME \cite{lime}}}
	\end{minipage}
	\begin{minipage}{2.9cm}
	\includegraphics[width=2.8cm]{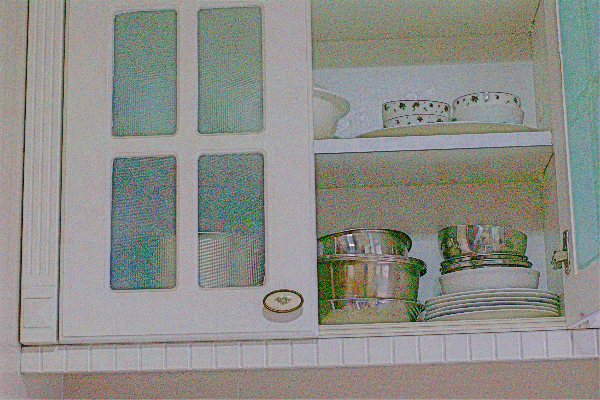}
	\vspace*{-2.5mm}
	\caption*{\small{17.78 / 0.492 / 0.430}}
	\vspace*{-4mm}
	\caption*{\small{RetinexNet \cite{retinexnet}}}
	\end{minipage}
	\begin{minipage}{2.9cm}
	\includegraphics[width=2.8cm]{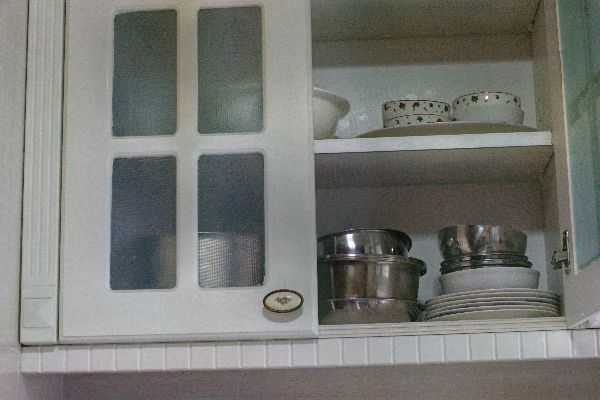}
	\vspace*{-2.5mm}
	\caption*{\small{15.20 / 0.737 / 0.250}}
	\vspace*{-4mm}
	\caption*{\small{EnlightenGAN \cite{enlightengan}}}
	\end{minipage}
	\begin{minipage}{2.9cm}
	\includegraphics[width=2.8cm]{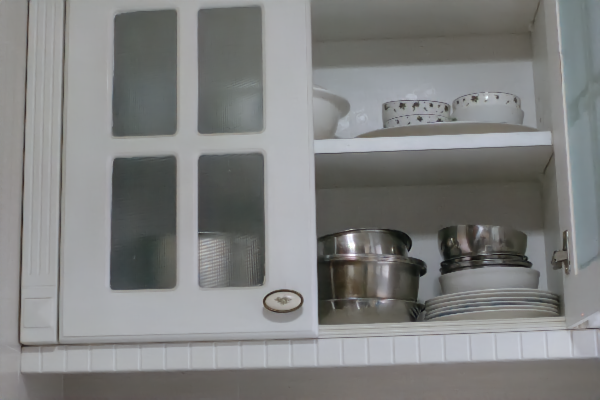}
	\vspace*{-2.5mm}
	\caption*{\small{17.72 / 0.896 / 0.121}}
	\vspace*{-4mm}
	\caption*{\small{DRBN \cite{DRBN}}}
	\end{minipage}
	}%
	\vspace*{-2mm}
	\quad
	\subfigure{
	\begin{minipage}{2.9cm}
	\includegraphics[width=2.8cm]{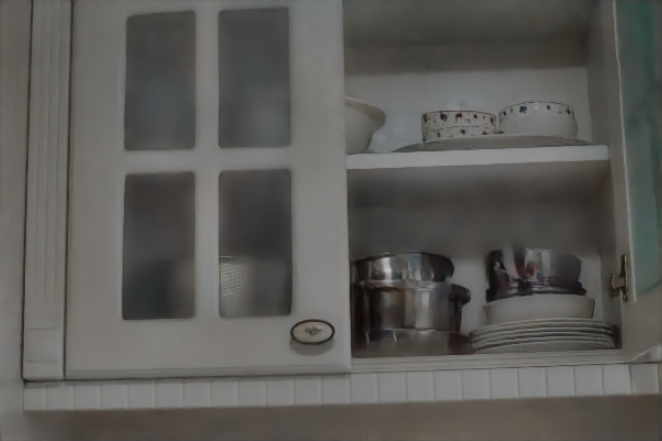}
	\vspace*{-2.5mm}
	\caption*{\small{10.94 / 0.717 / 0.300}}
	\vspace*{-4mm}
	\caption*{\small{KinD \cite{kind}}}
	\end{minipage}
	\begin{minipage}{2.9cm}
	\includegraphics[width=2.8cm]{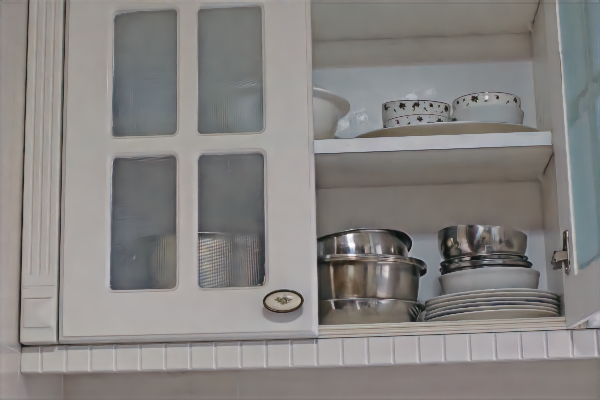}
	\vspace*{-2.5mm}
	\caption*{\small{17.71 / 0.882 / 0.144}}
	\vspace*{-4mm}
	\caption*{\small{KinD++ \cite{kind++}}}
	\end{minipage}
	\begin{minipage}{2.9cm}
	\includegraphics[width=2.8cm]{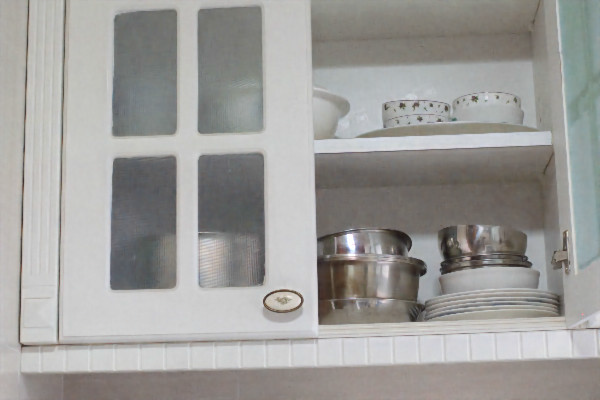}
	\vspace*{-2.5mm}
	\caption*{\small{\textbf{30.74} / 0.907 / 0.091}}
	\vspace*{-4mm}
	\caption*{\small{MIRNet \cite{01}}}
	\end{minipage}
	\begin{minipage}{2.9cm}
	\includegraphics[width=2.8cm]{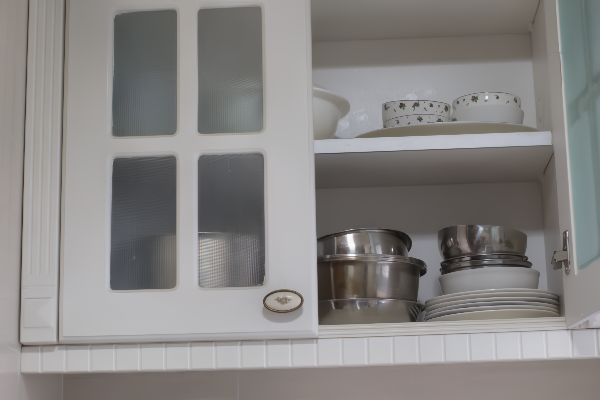}
	\vspace*{-2.5mm}
	\caption*{\small{19.33 / \underline{0.917} / \underline{0.086}}}
	\vspace*{-4mm}
	\caption*{\small{LLFlow \cite{LLFlow}}}
	\end{minipage}
	\begin{minipage}{2.9cm}
	\includegraphics[width=2.8cm]{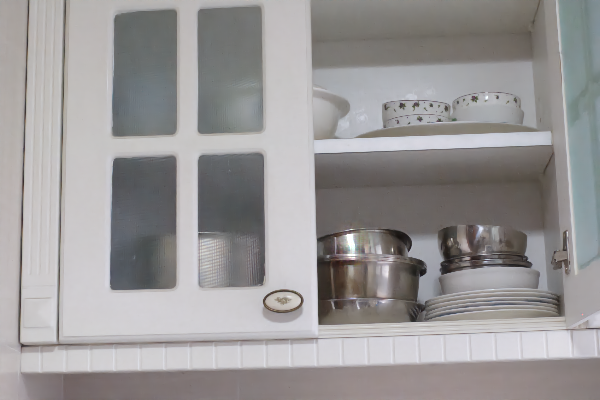}
	\vspace*{-2.5mm}
	\caption*{\small{\underline{28.34} / \textbf{0.928} / \textbf{0.076}}}
	\vspace*{-4mm}
	\caption*{\small{\textbf{HWMNet (Ours)}}}
	\end{minipage}
	\begin{minipage}{2.9cm}
	\includegraphics[width=2.8cm]{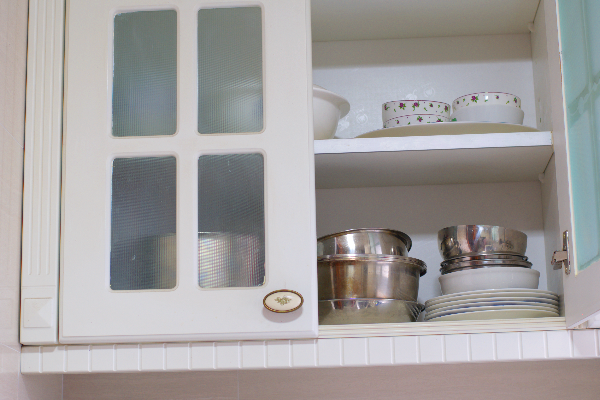}
	\vspace*{-2.5mm}
	\caption*{\small{PSNR / SSIM / LPIPS}}
	\vspace*{-4mm}
	\caption*{\small{GT}}
	\end{minipage}
	}%
	\vspace*{-4.5mm}  % --> 不知道有沒有規定不能縮這個距離 ˊ_>ˋ
\caption{Visual comparisons for low-light image enhancement on the LOL dataset \cite{lol}. The best and second best scores are \textbf{highlighted} and \underline{underlined}, respectively.}
\label{LOL_figure}
\end{figure*}

% --table-- of LOL dataset
\begin{table}[!b] \small
	\caption{Low-light image enhancement evaluation on the LOL dataset \cite{lol}. The best and second best scores are \textbf{highlighted} and \underline{underlined}, respectively.}
\vspace*{-4.5mm}  % --> 不知道有沒有規定不能縮這個距離 ˊ_>ˋ
	\label{LOL_table}
	\begin{center}
\setlength{\tabcolsep}{3.5mm}
	\begin{tabular}{l || ccc}
	\toprule[1.5 pt]
\multirow{2}{*}{\textbf{Methods}}&\multicolumn{3}{c}{Low-light dataset (LOL) \cite{lol}}\\
&PSNR $\uparrow$&SSIM $\uparrow$&LPIPS $\downarrow$\\
	\midrule[1.5 pt]
MSR \cite{MSR}  					&13.17&0.48&-\\
Zero-DCE \cite{zeroDCE}			&14.86&0.54&0.33\\
LIME \cite{lime}           			&16.76&0.56&0.35\\
Retinex-Net \cite{retinexnet}		&16.77&0.56&0.47\\
EnlightenGAN \cite{enlightengan}  &17.48&0.65&0.32\\
RUAS \cite{RUAS}					&18.23&0.72&0.35\\
GLAD \cite{gladnet}  				&19.72&0.70&-\\
DRBN \cite{DRBN}					&20.13&0.83&0.16\\
KinD \cite{kind}  					&20.87&0.80&0.17\\
KinD++ \cite{kind++}  				&21.30&0.82&0.16\\
MIRNet \cite{01}					&24.14&0.83&0.13\\
LLFlow \cite{LLFlow}				&\textbf{25.19}&\textbf{0.93}&\textbf{0.11}\\
	\midrule[1.5 pt]
\textbf{HWMNet (Ours)}          	&\underline{24.24}&\underline{0.85}&\underline{0.12}\\
	\bottomrule[1.5pt]
	\end{tabular}
	\end{center}
\end{table}

% --table-- of MIT-5K
\begin{table*}[!ht] \footnotesize
\caption{Low-light image enhancement evaluation on the MIT-FiveK dataset \cite{fivek}. The best and second best scores are \textbf{highlighted} and \underline{underlined}, respectively.}
\vspace*{-5mm}  % --> 不知道有沒有規定不能縮這個距離 ˊ_>ˋ
\label{fiveK_table}
\begin{center}
\setlength{\tabcolsep}{4.2mm}
\begin{tabular}{l cccccccc}
\toprule[1.5 pt]
Method&HDRNet \cite{HDR}&W-Box \cite{wbox}&DR \cite{dr}&DPE \cite{DPE}&DeepUPE \cite{DeepUPE}&MIRNet \cite{01}&HWMNet (Ours)\\
\midrule[1.5 pt]
PSNR&21.96&18.57&20.97&22.15&23.04&\underline{23.73}&\textbf{24.44}\\
SSIM&0.866&0.701&0.841&0.850&0.893&\textbf{0.925}&\underline{0.914}\\
\bottomrule[1.5pt]
\end{tabular}
\end{center}
\end{table*}

% --figure-- of visual result
\begin{figure}[!t] \footnotesize
\centering
	\subfigure{
	\begin{minipage}{2.8cm}
	\includegraphics[width=2.7cm]{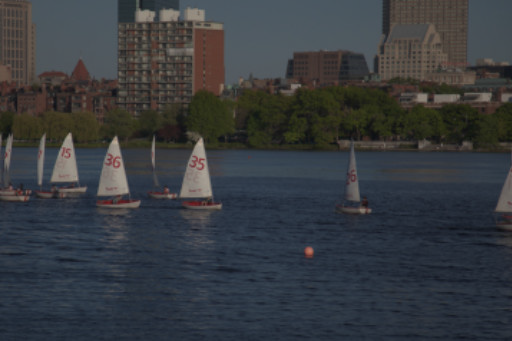}
	\vspace*{-2.5mm}
	\caption*{5.09/0.197/0.523}
	\vspace*{-4mm}
	\caption*{Low-light}
	\end{minipage}
	\begin{minipage}{2.8cm}
	\includegraphics[width=2.7cm]{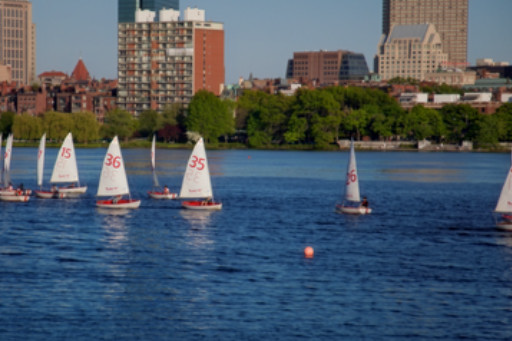}
	\vspace*{-2.5mm}
	\caption*{17.66/0.859/0.106}
	\vspace*{-4mm}
	\caption*{DPE \cite{DPE}}
	\end{minipage}
	\begin{minipage}{2.8cm}
	\includegraphics[width=2.7cm]{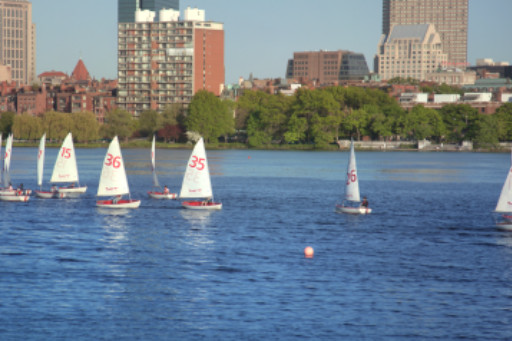}
	\vspace*{-2.5mm}
	\caption*{25.06/0.898/0.087}
	\vspace*{-4mm}
	\caption*{DeepUPE \cite{DeepUPE}}
	\end{minipage}
	}%
	\vspace*{-2mm}
	\quad
	\subfigure{
	\begin{minipage}{2.8cm}
	\includegraphics[width=2.7cm]{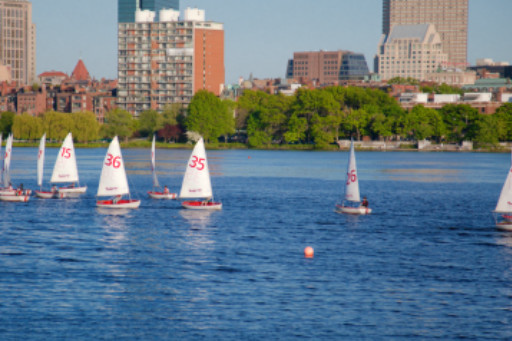}
	\vspace*{-2.5mm}
	\caption*{\textbf{31.07}/\underline{0.950}/\textbf{0.029}}
	\vspace*{-4mm}
	\caption*{MIRNet \cite{01}}
	\end{minipage}
	\begin{minipage}{2.8cm}
	\includegraphics[width=2.7cm]{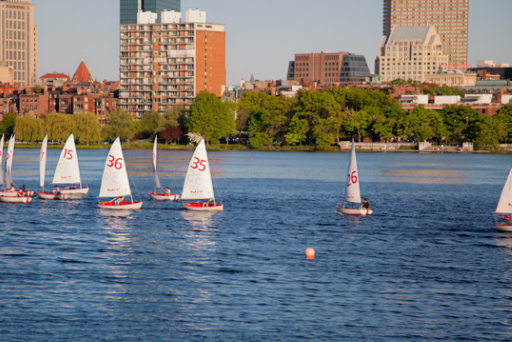}
	\vspace*{-2.5mm}
	\caption*{\underline{26.51}/\textbf{0.964}/\underline{0.032}}
	\vspace*{-4mm}
	\caption*{\textbf{HWMNet (Ours)}}
	\end{minipage}
	\begin{minipage}{2.8cm}
	\includegraphics[width=2.7cm]{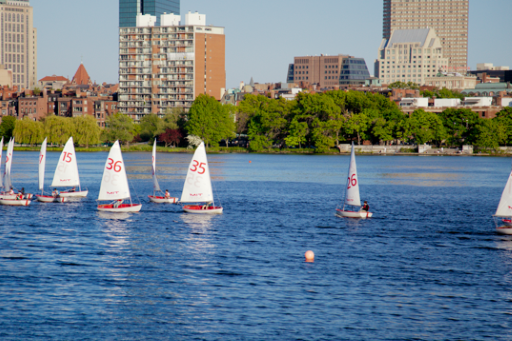}
	\vspace*{-2.5mm}
	\caption*{PSNR/SSIM/LPIPS}
	\vspace*{-4mm}
	\caption*{GT}
	\end{minipage}
	}%
	%\vspace*{-4.5mm}  % --> 不知道有沒有規定不能縮這個距離 ˊ_>ˋ
\caption{Visual comparisons for low-light image enhancement on the MIT-Adobe FiveK dataset \cite{fivek}.}% Best and second best scores are \textbf{highlighted} and \underline{underline}, respectively.
\label{fiveK_figure}
\end{figure}

\subsection{Experiment Datasets}
In this section, we will introduce two datasets for real-world low-light image enhancement. In the following experiments, we train separate models for different datasets. And the training, testing set details are illustrated below.

\noindent\textbf{LOL.} Low-light dataset \cite{lol} provides 485 and 15 images for training and testing, respectively. All the image sizes are $600 \times 400$, and each image pair in LOL dataset consists of a low-light image and its corresponding normal brightness ground-truth. In the experiments of LOL dataset, we randomly crop each training image to 10 $256 \times 256$ patches. As for the validation set, we use center cropped $256 \times 256$ patch for each testing image. 

\noindent\textbf{MIT-Adobe FiveK.} To train another dataset, MIT-Adobe FiveK \cite{fivek} contains 5000 images including both indoor and outdoor scenes captured with single-lens reflex (SLR) cameras in different lighting conditions. We follow the same experimental setup as other image enhancement methods \cite{01} which consider the enhanced images of expert C (one of the  experts from A to E) as the ground-truth images. Furthermore, the first 4500 images are for training and the last 500 images are for testing. 
%However, because this dataset is all in RAW format; that is, all the information recorded by the camera sensor is preserved, and can easily export images with any resolution by the application.
 
%In the experiment of MIT-Adobe FiveK dataset, we random crop $256 \times 256$ each training image to 8 patches. As for the validation set, we use center crop $256 \times 256$ each testing image same as the training setting of LOL. 

\subsection{Image Enhancement Performance}
\noindent\textbf{Evaluation on LOL.} In Table~\ref{LOL_table} and Fig.~\ref{LOL_figure}, we compare our HWMNet with the prior-based method (e.g., MSR \cite{MSR}) and other competitive enhancement learning-based methods on LOL dataset \cite{lol}. According to Table~\ref{LOL_table}, we could observe the proposed HWMNet achieves runner-up performance for all metrics (PSNR, SSIM and LPIPS) on LOL testing dataset, which means that our enhanced images are more perceptually faithful. Also, among the three top-performed methods (MIRNet \cite{01}, LLFlow \cite{LLFlow}, and our HWMNet), the floating-point operations per second (FLOPs) for MIRNet, LLFlow and HWMNet are 2.84T, 1.05T and 0.92T (tested on $400 \times 592$ color image, where T is ${10}^{12}$), respectively. This means the proposed HWMNet has the least computational complexity among the three best-performed models. We think it is attributed to the design of HWAB, where half of the features passing through the attention can decrease the computational complexity \cite{hinet}, and the design of wavelet attention could gain more semantic information and details in the training process to keep satisfactory results.
%1) Especially the visual image results between MIRNet \cite{01} and ours method in Fig.~\ref{LOL_figure}, it can be observed that the enhanced image of MIRNet has the best PSNR values, but there is a obvious speckle or noise on the restored image.

\noindent\textbf{Evaluation on MIT-Adobe FiveK.} As for the MIT-5K \cite{fivek}, the average performance scores on the testing set and comparisons of some enhanced image are shown in Table~\ref{fiveK_table} and Fig.~\ref{fiveK_figure}. It shows that the proposed HWMNet still achieves state-of-the-art performance on both PSNR and SSIM. 

%Actually, the enhanced performances are good or not, always depends on the subjective feeling. You can easily upload personal images to the following website to view the restored results: \\ \url{https://reurl.cc/OpVdOX}.

% --section-- of Conclusion
\section{Conclusion}
In this paper, we present a restoration model HWMNet and achieve the competitive performances on low-light image enhancements. We improved the hierarchical model M-Net which is originally proposed for medical image segmentation. The novel design of M-Net+ has the advantage of enriching features with different resolutions. Moreover, the proposed HWAB focuses on the features of wavelet domain which can enrich the semantic information, too. In the future work, we are going to improve current model to handle different restoration tasks such as image denoising and deblurring.

\bibliographystyle{IEEEbib}
\fontsize{8.5pt}{8.5pt}
\selectfont
\bibliography{refs}

\begin{thebibliography}{10}

\bibitem{01}
Syed~Waqas Zamir, Aditya Arora, Salman Khan, Munawar Hayat, Fahad~Shahbaz Khan,
  Ming-Hsuan Yang, and Ling Shao,
\newblock ``Learning enriched features for real image restoration and
  enhancement,''
\newblock in {\em Computer Vision--ECCV 2020: 16th European Conference,
  Glasgow, UK, August 23--28, 2020, Proceedings, Part XXV 16}. Springer, 2020,
  pp. 492--511.

\bibitem{adaptive}
Stephen~M Pizer, E~Philip Amburn, John~D Austin, Robert Cromartie, Ari
  Geselowitz, Trey Greer, Bart ter Haar~Romeny, John~B Zimmerman, and Karel
  Zuiderveld,
\newblock ``Adaptive histogram equalization and its variations,''
\newblock {\em Computer vision, graphics, and image processing}, vol. 39, no.
  3, pp. 355--368, 1987.

\bibitem{contrast}
Yeong-Taeg Kim,
\newblock ``Contrast enhancement using brightness preserving bi-histogram
  equalization,''
\newblock {\em IEEE transactions on Consumer Electronics}, vol. 43, no. 1, pp.
  1--8, 1997.

\bibitem{MSR}
Zia-ur Rahman, Daniel~J Jobson, and Glenn~A Woodell,
\newblock ``Multi-scale retinex for color image enhancement,''
\newblock in {\em Proceedings of 3rd IEEE International Conference on Image
  Processing}. IEEE, 1996, vol.~3, pp. 1003--1006.

\bibitem{DCP}
Kaiming He, Jian Sun, and Xiaoou Tang,
\newblock ``Single image haze removal using dark channel prior,''
\newblock {\em IEEE transactions on pattern analysis and machine intelligence},
  vol. 33, no. 12, pp. 2341--2353, 2010.

\bibitem{2019ntire}
Andrey Ignatov and Radu Timofte,
\newblock ``Ntire 2019 challenge on image enhancement: Methods and results,''
\newblock in {\em Proceedings of the IEEE/CVF Conference on Computer Vision and
  Pattern Recognition Workshops}, 2019, pp. 0--0.

\bibitem{kind}
Qiming Zhang, Yao He, Nan Luo, Shashank~J Patel, Yanjie Han, Ranran Gao,
  Madhura Modak, Sebastian Carotta, Christian Haslinger, David Kind, et~al.,
\newblock ``Landscape and dynamics of single immune cells in hepatocellular
  carcinoma,''
\newblock {\em Cell}, vol. 179, no. 4, pp. 829--845, 2019.

\bibitem{kind++}
Yonghua Zhang, Xiaojie Guo, Jiayi Ma, Wei Liu, and Jiawan Zhang,
\newblock ``Beyond brightening low-light images,''
\newblock {\em International Journal of Computer Vision}, vol. 129, no. 4, pp.
  1013--1037, 2021.

\bibitem{RUAS}
Risheng Liu, Long Ma, Jiaao Zhang, Xin Fan, and Zhongxuan Luo,
\newblock ``Retinex-inspired unrolling with cooperative prior architecture
  search for low-light image enhancement,''
\newblock in {\em Proceedings of the IEEE/CVF Conference on Computer Vision and
  Pattern Recognition}, 2021, pp. 10561--10570.

\bibitem{LLFlow}
Yufei Wang, Renjie Wan, Wenhan Yang, Haoliang Li, Lap-Pui Chau, and Alex~C Kot,
\newblock ``Low-light image enhancement with normalizing flow,''
\newblock {\em arXiv preprint arXiv:2109.05923}, 2021.

\bibitem{mnet}
Raghav Mehta and Jayanthi Sivaswamy,
\newblock ``M-net: A convolutional neural network for deep brain structure
  segmentation,''
\newblock in {\em 2017 IEEE 14th International Symposium on Biomedical Imaging
  (ISBI 2017)}. IEEE, 2017, pp. 437--440.

\bibitem{unet}
Olaf Ronneberger, Philipp Fischer, and Thomas Brox,
\newblock ``U-net: Convolutional networks for biomedical image segmentation,''
\newblock in {\em International Conference on Medical image computing and
  computer-assisted intervention}. Springer, 2015, pp. 234--241.

\bibitem{fpd}
Sukesh Adiga and Jayanthi Sivaswamy,
\newblock ``Fpd-m-net: Fingerprint image denoising and inpainting using m-net
  based convolutional neural networks,''
\newblock in {\em Inpainting and Denoising Challenges}, pp. 51--61. Springer,
  2019.

\bibitem{RDN}
Yulun Zhang, Yapeng Tian, Yu~Kong, Bineng Zhong, and Yun Fu,
\newblock ``Residual dense network for image super-resolution,''
\newblock in {\em Proceedings of the IEEE conference on computer vision and
  pattern recognition}, 2018, pp. 2472--2481.

\bibitem{cycleisp}
Syed~Waqas Zamir, Aditya Arora, Salman Khan, Munawar Hayat, Fahad~Shahbaz Khan,
  Ming-Hsuan Yang, and Ling Shao,
\newblock ``Cycleisp: Real image restoration via improved data synthesis,''
\newblock in {\em Proceedings of the IEEE/CVF Conference on Computer Vision and
  Pattern Recognition}, 2020, pp. 2696--2705.

\bibitem{SENet}
Jie Hu, Li~Shen, and Gang Sun,
\newblock ``Squeeze-and-excitation networks,''
\newblock in {\em Proceedings of the IEEE conference on computer vision and
  pattern recognition}, 2018, pp. 7132--7141.

\bibitem{hinet}
Liangyu Chen, Xin Lu, Jie Zhang, Xiaojie Chu, and Chengpeng Chen,
\newblock ``Hinet: Half instance normalization network for image restoration,''
\newblock in {\em Proceedings of the IEEE/CVF Conference on Computer Vision and
  Pattern Recognition}, 2021, pp. 182--192.

\bibitem{LPIPS}
Richard Zhang, Phillip Isola, Alexei~A Efros, Eli Shechtman, and Oliver Wang,
\newblock ``The unreasonable effectiveness of deep features as a perceptual
  metric,''
\newblock in {\em Proceedings of the IEEE conference on computer vision and
  pattern recognition}, 2018, pp. 586--595.

\bibitem{zeroDCE}
Chunle Guo, Chongyi Li, Jichang Guo, Chen~Change Loy, Junhui Hou, Sam Kwong,
  and Runmin Cong,
\newblock ``Zero-reference deep curve estimation for low-light image
  enhancement,''
\newblock in {\em Proceedings of the IEEE/CVF Conference on Computer Vision and
  Pattern Recognition}, 2020, pp. 1780--1789.

\bibitem{lime}
Xiaojie Guo, Yu~Li, and Haibin Ling,
\newblock ``Lime: Low-light image enhancement via illumination map
  estimation,''
\newblock {\em IEEE Transactions on image processing}, vol. 26, no. 2, pp.
  982--993, 2016.

\bibitem{retinexnet}
Chen Wei, Wenjing Wang, Wenhan Yang, and Jiaying Liu,
\newblock ``Deep retinex decomposition for low-light enhancement,''
\newblock {\em arXiv preprint arXiv:1808.04560}, 2018.

\bibitem{enlightengan}
Yifan Jiang, Xinyu Gong, Ding Liu, Yu~Cheng, Chen Fang, Xiaohui Shen, Jianchao
  Yang, Pan Zhou, and Zhangyang Wang,
\newblock ``Enlightengan: Deep light enhancement without paired supervision,''
\newblock {\em IEEE Transactions on Image Processing}, vol. 30, pp. 2340--2349,
  2021.

\bibitem{DRBN}
Wenhan Yang, Shiqi Wang, Yuming Fang, Yue Wang, and Jiaying Liu,
\newblock ``From fidelity to perceptual quality: A semi-supervised approach for
  low-light image enhancement,''
\newblock in {\em Proceedings of the IEEE/CVF Conference on Computer Vision and
  Pattern Recognition}, 2020, pp. 3063--3072.

\bibitem{lol}
Chen Wei, Wenjing Wang, Wenhan Yang, and Jiaying Liu,
\newblock ``Deep retinex decomposition for low-light enhancement,''
\newblock {\em arXiv preprint arXiv:1808.04560}, 2018.

\bibitem{gladnet}
Wenjing Wang, Chen Wei, Wenhan Yang, and Jiaying Liu,
\newblock ``Gladnet: Low-light enhancement network with global awareness,''
\newblock in {\em 2018 13th IEEE International Conference on Automatic Face \&
  Gesture Recognition (FG 2018)}. IEEE, 2018, pp. 751--755.

\bibitem{fivek}
Vladimir Bychkovsky, Sylvain Paris, Eric Chan, and Fr{\'e}do Durand,
\newblock ``Learning photographic global tonal adjustment with a database of
  input/output image pairs,''
\newblock in {\em CVPR 2011}. IEEE, 2011, pp. 97--104.

\bibitem{HDR}
Micha{\"e}l Gharbi, Jiawen Chen, Jonathan~T Barron, Samuel~W Hasinoff, and
  Fr{\'e}do Durand,
\newblock ``Deep bilateral learning for real-time image enhancement,''
\newblock {\em ACM Transactions on Graphics (TOG)}, vol. 36, no. 4, pp. 1--12,
  2017.

\bibitem{wbox}
Yuanming Hu, Hao He, Chenxi Xu, Baoyuan Wang, and Stephen Lin,
\newblock ``Exposure: A white-box photo post-processing framework,''
\newblock {\em ACM Transactions on Graphics (TOG)}, vol. 37, no. 2, pp. 1--17,
  2018.

\bibitem{dr}
Jongchan Park, Joon-Young Lee, Donggeun Yoo, and In~So Kweon,
\newblock ``Distort-and-recover: Color enhancement using deep reinforcement
  learning,''
\newblock in {\em Proceedings of the IEEE Conference on computer vision and
  pattern recognition}, 2018, pp. 5928--5936.

\bibitem{DPE}
Yu-Sheng Chen, Yu-Ching Wang, Man-Hsin Kao, and Yung-Yu Chuang,
\newblock ``Deep photo enhancer: Unpaired learning for image enhancement from
  photographs with gans,''
\newblock in {\em Proceedings of the IEEE Conference on Computer Vision and
  Pattern Recognition}, 2018, pp. 6306--6314.

\bibitem{DeepUPE}
Ruixing Wang, Qing Zhang, Chi-Wing Fu, Xiaoyong Shen, Wei-Shi Zheng, and Jiaya
  Jia,
\newblock ``Underexposed photo enhancement using deep illumination
  estimation,''
\newblock in {\em Proceedings of the IEEE/CVF Conference on Computer Vision and
  Pattern Recognition}, 2019, pp. 6849--6857.

\end{thebibliography}

\end{document}